\documentclass[journal]{IEEEtran}

\IEEEoverridecommandlockouts

\usepackage{cite}
\usepackage{amsmath,amssymb,amsfonts}
\usepackage{algorithmic}
\usepackage{graphicx}
\usepackage{textcomp}
\usepackage{subcaption}
\usepackage{changepage}
\usepackage[font=small,labelfont=bf]{caption}
\usepackage{caption}

\usepackage{xcolor}
\def\BibTeX{{\rm B\kern-.05em{\sc i\kern-.025em b}\kern-.08em
    T\kern-.1667em\lower.7ex\hbox{E}\kern-.125emX}}
\usepackage{blindtext} 

\usepackage[left=0.68in,right=0.673in,top=0.75in,bottom=1.1in]{geometry}

\begin{document}

\title{ Multi-Layer Network Formation through HAPS Base Station and Transmissive RIS-Equipped UAV  }


\author{\IEEEauthorblockN{Faical Khennoufa$^{1}$,
Khelil Abdellatif$^{1}$,
Halim Yanikomeroglu$^{2}$,  
Metin Ozturk$^{2,3}$,\\ Taissir Elganimi$^{4}$, 
Ferdi Kara$^{5,6}$, and Khaled Rabie$^{7}$}

\IEEEauthorblockA{$^{1}$LGEERE Lab, Department of Electrical Engineering, El-Oued University, El-Oued,  Algeria}

\IEEEauthorblockA{$^{2}$NTN Lab, Department of Systems and Computer Engineering, Carleton University, Ottawa, Canada}

\IEEEauthorblockA{$^{3}$Electrical and Electronics Engineering, Ankara Yıldırım Beyazıt University, Ankara, Türkiye}

\IEEEauthorblockA{$^{4}$Department of Electrical and Electronic Engineering, University of Tripoli, Libya}

\IEEEauthorblockA{$^{5}$Division of Communications Systems, KTH Royal Institute of Technology, Sweden}

\IEEEauthorblockA{$^{6}$Department of Computer Engineering, Zonguldak Bulent Ecevit University, Zonguldak, Türkiye}

\IEEEauthorblockA{$^{7}$Department of Computer Engineering \& Center for Communication Systems and Sensing \\ at King Fahd University of Petroleum \& Minerals (KFUPM), Dhahran, Saudi Arabia}

}   








\markboth{IEEE Wireless Communications and Networking Conference (WCNC) 2025, Milan, Italy}%
{Shell \MakeLowercase{\textit{et al.}}: A Sample Article Using IEEEtran.cls for IEEE Journals}

\maketitle
\begin{abstract}
In order to bolster future wireless networks, there has been a great deal of interest in non-terrestrial networks, especially aerial platforms including high-altitude platform stations (HAPS) and uncrewed aerial vehicles (UAVs). These platforms can integrate advanced technologies such as reconfigurable intelligent surfaces (RIS) and non-orthogonal multiple access (NOMA). In this regard, this paper proposes a multi-layer network architecture consisting of HAPS and UAV, where the former acts as a HAPS super macro base station (HAPS-SMBS), while the latter serves as a relay node for the ground Internet of Things (IoT) devices. The UAV is equipped with active transmissive RIS, which is a novel technology with promising benefits.
We also utilize multiple-input single-output (MISO) technology, i.e., multiple antennas at the HAPS-SMBS and a single antenna at the IoT devices. Additionally, we consider NOMA as the multiple access technology as well as the existence of hardware impairments as a practical limitation. 
We compare the proposed system model with various scenarios, all involving the HAPS-SMBS and RIS-equipped UAV relay combination, but with different types of RIS, antenna configurations, and access technologies.
Sum rate and energy efficiency are used as performance metrics, and the findings demonstrate that, in comparison to all benchmarks, the proposed system yields significant performance gains. Moreover, hardware impairment limits the system performance at high transmit power levels. 
\end{abstract}

\begin{IEEEkeywords}
HAPS, non-terrestrial networks, hardware impairment, MISO, NOMA, and transmissive RIS.
\end{IEEEkeywords}

\section{Introduction}
The sixth-generation (6G) technology standard for cellular networks entails significant changes in network structures and technologies. 
In this regard, non-terrestrial networks (NTN) integrated with 6G offer a cost-effective approach to establishing seamless and expansive wireless connectivity across diverse environments—including rural, remote, and urban areas \cite{alfattani2021link}.
According to the Third Generation Partnership Project (3GPP), NTN includes three main kinds of flying objects, namely satellites, high altitude platform stations (HAPS), and uncrewed aerial vehicles (UAV) \cite{3gpp2018study}. NTN can act as an extension to the current ground networks, increasing their capacity and extending coverage, which aligns with the usage scenarios and capabilities described by the International Telecommunication Union's (ITU) vision document for International Mobile Telecommunications (IMT)-2030 {\color{black}(i.e., 6G)}~\cite{ITU2030}. 

As another promising technology, smart radio environments, named reconfigurable intelligent surfaces (RIS), have recently been proposed for intelligently controlling wireless channels to achieve improved communication performance~\cite{zhang2022active}. 
{\color{black}In particular, a RIS array consists of multiple passive elements that reflect electromagnetic waves in a controlled manner, altering signal propagation in a wireless environment.}
RIS can be mounted on aerial platforms as a reflective layer between {\color{black}the ground/aerial} stations and ground devices~\cite{alfattani2021link}.
In the absence of direct links, the joint optimization of trajectory and phase shift for a HAPS base station (BS) with RIS-equipped UAV (UAV-RIS) for a multiple-input single-output (MISO) system was investigated in~\cite{gao2021aerial}. 
The authors in~\cite{nguyen2022design} obtained the outage probability, average transmission rate, and spectral efficiency for a satellite-assisted HAPS with UAV-RIS, when both act as relays.  
{\color{black} Moreover, one of the key challenges facing NTNs is the efficient use of the limited available spectrum, making designing the right multiple access technology critical to improving network capacity. Compared to orthogonal multiple access (OMA), non-orthogonal multiple access (NOMA) addresses this challenge by allowing multiple users to share the same frequency bands without requiring additional bandwidth, thereby enhancing the performance and capacity of NTNs in the upcoming 6G landscape~\cite{beddiaf2022unified,khennoufa2024error}. }
On the other hand, the active RIS with NOMA and hardware impairments (HWI) were considered to maximize the sum rate in the absence of aerial platforms in~\cite{yang2023sum,yue2024exploiting}. 


{\color{black}The majority of the} previous works, as in \cite{gao2021aerial}, take into account two dimensions (2D) for reflective RIS, meaning that 2D reflective RIS will be directed in one direction to serve devices at the expense of others. 
{\color{black}Based on the structure of the aerial platforms, using a 2D reflective RIS installed on a UAV may not be an effective solution because it only covers one side (i.e., 180 degrees). To solve this issue, transmissive RIS (TRIS) was introduced in~\cite{zeng2021reconfigurable}, where the transmitted signal can completely penetrate the structure of RIS.
Moreover, the works in the literature predominantly consider a passive RIS technology in their investigation~\cite{gao2021aerial,nguyen2022design}. Nevertheless, there are some disadvantages associated with the passive RIS. For example, the received signal experiences product/double path-loss attenuation (i.e., the transmitter to RIS and RIS to receiver paths) in the absence of signal amplification, causing it to become dramatically low, thereby the potential of RIS is significantly restricted by this ``double path-loss" attenuation. 
In this regard, the active RIS that has power amplification capabilities has been recommended to resolve this issue~\cite{zeng2021reconfigurable}.}

{\color{black} 
HAPS, with its wide-area coverage and high capacity, functions as a permanent communications node in the stratosphere to reach unserved areas, and it can also support terrestrial networks to enhance coverage. UAVs are adaptable, low-altitude nodes that address problems with mobility and coverage.
The TRIS onboard UAV ensures better signal propagation everywhere, while active TRIS technology reduces dual path loss.
The NOMA technology is used for its ability to serve multiple devices simultaneously using the same frequency bands, which improves spectrum efficiency, while the MISO technology helps reduce interference and improve signal quality, which improves overall system performance.}
To the best of the authors' knowledge, the literature has not considered active TRIS with aerial platforms using technologies including HAPS, UAV, MISO, and NOMA. 
{\color{black}In order to achieve an integrated and robust scheme that can meet the challenges of NTN and improve the overall system performance, this paper investigates} the passive/active TRIS for aerial platforms when the HAPS acts as a BS, referred to as HAPS super macro BS (HAPS-SMBS) (or HAPS as an IMT BS (HIBS) as referred in ITU documents)~\cite{ITU2023}. In particular, TRIS is mounted on a UAV (UAV-TRIS) to serve as a relay node for the Internet of things (IoT) ground devices.  Additionally, a MISO system is taken into account in our system, where the transmission antenna selection is used. For a more practical scenario, we also consider the effect of HWI {\color{black}and channel state information (CSI).} 
The contributions of this paper are as follows: 

\begin{itemize}
    \item We introduce the application of passive/active UAV-TRIS for NOMA networks in the presence of HWI when HAPS is employed as an SMBS to serve multiple IoT ground devices. We consider the transmission antenna selection (TAS) system, where the HAPS-SMBS has multiple antennas while ground IoT devices have a single antenna. 
    \item The effect of HWI {\color{black}and imperfect CSI} on the system performance is taken into consideration {\color{black}in order to make the analysis more practical.} 
    \item The sum rate and energy efficiency of the active/passive UAV-TRIS with TAS and HAPS-SMBS are obtained under the effect of HWI. 
    As a benchmark, our system is compared to three schemes as follows. The first scheme is a HAPS-SMBS-assisted UAV relay equipped with active TRIS and a single-input single-output system. The second scheme is HAPS-SMBS-assisted UAV equipped with amplify-and-forward (AF) relaying supported by TAS. The third scheme is HAPS-SMBS-assisted UAV equipped with passive TRIS supported by TAS.
    

\end{itemize}

\begin{figure}[]
    \centering
      \includegraphics[width=6.5cm,height=6.5cm]{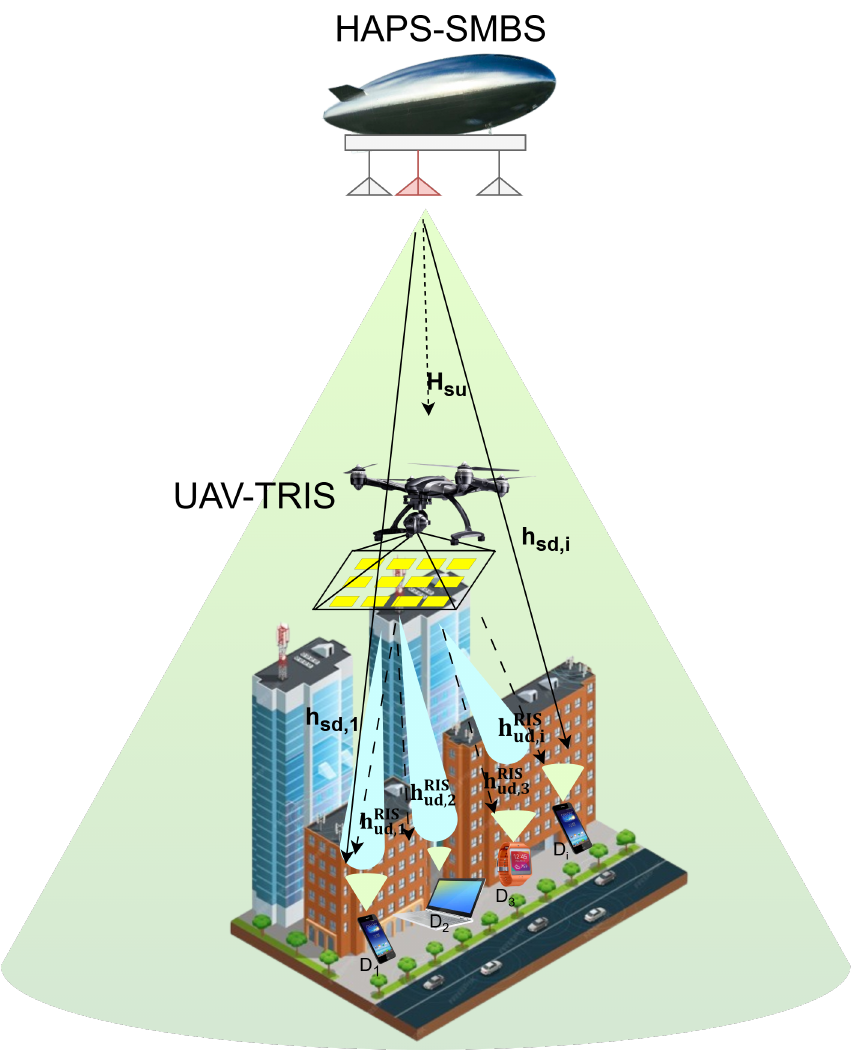}
    \caption{HAPS-SMBS assisted UAV-TRIS scheme for IoT NOMA networks.}
    \label{constellations}
\end{figure}

\section{System Model}

As presented in Fig. 1, our proposed system consists of a single HAPS-SMBS with $M$ antennas, a single UAV, TRIS with $N$ elements, and $L$ IoT ground devices. We consider the TRIS to be deployed on the UAV, and NOMA is used as a multiple-access technology to serve the IoT ground devices. 
 It is assumed that the HAPS-SMBS acts as a transmitter (i.e., BS), and the UAV-TRIS operates as a relay. IoT ground devices receive two signals: A direct signal from the HAPS-SMBS and an indirect signal through the UAV-TRIS. We consider practical limitations (e.g., HWI) that exist in the HAPS-SMBS and IoT ground devices. {\color{black}It is assumed that the CSI is available at all the nodes. Precisely, we consider that the channel of the direct link (i.e., HAPS-SMBS to IoT ground devices) is imperfect due to high attenuation or blockages, leading to higher estimation errors, while assuming perfect CSI for the RIS links (i.e., HAPS-SMBS to UAV and UAV to IoT ground devices) due to better estimation techniques~\cite{beddiaf2022unified,khennoufa2024error}}. The details of the system model are provided in the following paragraphs.

\subsection{HAPS-SMBS and UAV Relaying without TRIS}
We consider that in addition to the direct link between the HAPS-SMBS and IoT ground devices, there is also an indirect link through the UAV. The UAV operates as a relay node, which uses the AF protocol, and we consider the UAV to have a single receiver and transmitter antenna. The HAPS-SMBS communicates with the IoT ground devices using NOMA; thus, the HAPS-SMBS transmits a superimposed coding signal with different power allocation coefficients according to channel gain (i.e., channel gain for direct links $|\mathbf{h}_{\text{sd},1}|^2< |\mathbf{h}_{\text{sd},2}|^2< ...<|\mathbf{h}_{\text{sd},i}|^2$ and channel gain for indirect links $|\mathbf{h}_{\text{su}}^{\text{AF}}\mathbf{h}_{\text{ud},1}^{\text{AF}}|^2< |\mathbf{h}_{\text{su}}^{\text{AF}}\mathbf{h}_{\text{ud},2}^{\text{AF}}|^2< ...<|\mathbf{h}_{\text{su}}^{\text{AF}}\mathbf{h}_{\text{ud},i}^{\text{AF}}|^2$) to IoT ground devices directly and through the UAV relaying. The UAV amplifies the received signal and forwards it to the IoT ground devices using its power. Thus, the received signal through direct and indirect links at IoT ground devices is given by
 \begin{equation} \label{eq:4}
 \begin{split}
& y_{\text{AF},i}= \mathit{h}_{\text{ud},i}^{\text{AF}}(\Lambda  \sqrt{P_{\text{u}}} (\mathbf{h}_{\text{su}}^{\text{AF}} (\sqrt{P_{\text{t}}}x_{\text{sc}}+\eta_{\text{t,su}}) +\eta_{\text{r,su}} + \aleph)
  + \eta_{\text{t,ud},i})\\& \ \ \ \ \ \ \ \ +\eta_{\text{r,ud},i}  
   + \mathbf{h}_{\text{sd},i} (\sqrt{P_{\text{t}}}x_{\text{sc}}+\eta_{\text{t,sd},i}) +\eta_{\text{r,sd},i} + \aleph,
 \end{split}
\end{equation}
where
$\Lambda  =\sqrt{\frac{1}{|\mathbf{h}_{\text{su}}^{\text{AF}}|^2 P_{\text{t}}+\sigma^2}}$,
{\color{black}$\mathbf{h}_{\text{sd},i}=\hat{h}_{\text{sd},i}+ \Delta\mathbf{h}_{\text{sd},i}$, $\hat{h}_{\text{sd},i}$ is the estimation channel of HAPS-SMBS to IoT ground devices, $\Delta\mathbf{h}_{\text{sd},i}$ is the estimation error}.
$x_{\text{sc}}=\sum_{i=1}^{L} \alpha_{i} s_{i}$, $\alpha_{i}$ is the power allocation coefficients, in which $\alpha_{1}>\alpha_{2}>...>\alpha_{i}$ and $ \sum_i^L\alpha_{i}=1$, $s_{i}$ is the signals of IoT ground devices, $P_{\text{t}}$ and $P_{\text{u}}$ are the transmit power of HAPS-SMBS and UAV, respectively.
$\aleph$ is the additive white Gaussian noise (AWGN) which follows ${\aleph} \sim \mathcal{CN}(0,\ \sigma^2)$.~$\eta_{\text{t,su}}$, $\eta_{\text{t,ud,i}}$, $\eta_{\text{t,sd},i}$, $\eta_{\text{r,su}}$, $\eta_{\text{r,ud},i}$, and $\eta_{\text{r,sd},i}$ are the independent distortion noise terms defined as \cite{beddiaf2022unified,khennoufa2024error}
\begin{math}
 \mathbf {\eta}_{\text{t,su}} \sim \mathcal{CN}(0,\ {k}_{\text{t,su}}^2 {P}_{\text{t}}), 
 \end{math}
\begin{math}
 \mathbf {\eta}_{\text{t,ud},i} \sim \mathcal{CN}(0, {k}_{\text{t,ud},i}^2 {P}_{\text{u}}),
 \end{math}
 \begin{math}
 \mathbf {\eta}_{\text{t,sd},i} \sim \mathcal{CN}(0,\ {k}_{\text{t,sd},i}^2 {P}_{\text{t}}), 
 \end{math}
\begin{math}
 \mathbf {\eta}_{\text{r,su}} \sim \mathcal{CN}(0,{k}_{\text{r,su}}^2 {P}_{\text{t}} |\mathbf{h}_{\text{su}}^{\text{AF}}|^2), 
\end{math}
 \begin{math}
 \mathbf {\eta}_{\text{r,ud},i} \sim \mathcal{CN}(0, {k}_{\text{r,ud},i}^2 {P}_{\text{u}} |\mathit{h}_{\text{ud},i}^{\text{AF}}|^2) 
 , 
\end{math}
 and
{\color{black} \begin{math}\mathbf {\eta}_{\text{r,sd},i} \sim \mathcal{CN}(0,{k}_{\text{r,sd},i}^2 {P}_{\text{t}} |\hat{h}_{\text{sd},i}|^2),
\end{math}}
in which $k_{\text{t,su}}$, $k_{\text{t,ud},i}$, $k_{\text{t,sd},i}$, $k_{\text{r,su}}$, $k_{\text{r,ud},i}$, and $k_{\text{r,sd},i}$ represent levels of impairment at the transceiver, respectively. According to \cite{beddiaf2022unified,khennoufa2024error}, the impact of transceiver HWI can be measured by the aggregate level of impairments, $K_{\text{su}}^2={k}_{\text{t,su}}^2 + {k}_{\text{r,su}}^2$, $K_{\text{ud},i}^2={k}_{\text{t,ud},i}^2 + {k}_{\text{r,ud},i}^2$, and $K_{\text{sd},i}^2={k}_{\text{t,sd},i}^2 + {k}_{\text{r,sd},i}^2$.
The Rician distribution channel vector between the HAPS-SMBS and UAV, denoted by $\mathbf{h}_{\text{su}}^{\text{AF}}  \in \mathbb{C}^{1 \times M}$, can be given as
 \begin{equation} \label{eq:4}
\mathbf{h}_{\text{su}}^{\text{AF}}=   \sqrt{\frac{1}{\mathfrak{Q}_{\text{su}}} \frac{Z_{\text{su}}}{Z_{\text{su}}+1}} \bar{\mathbf{h}}_{\text{su}}^{\text{AF}}+\sqrt{\frac{1}{\mathfrak{Q}_{\text{su}}} \frac{1}{Z_{\text{su}}+1}} \tilde{\mathbf{h}}_{\text{su}}^{\text{AF}},
\end{equation}
while the Rician distribution channel vector between the UAV and IoT ground devices, denoted by $\mathit{h}_{\text{ud},i}^{\text{AF}}  \in \mathbb{C}^{1 \times 1}$, is as follows
 \begin{equation} \label{eq:4}
\mathit{h}_{\text{ud},i}^{\text{AF}}=   \sqrt{\frac{1}{\mathfrak{Q}_{\text{ud},i}} \frac{Z_{\text{ud},i}}{Z_{\text{ud},i}+1}} \bar{\mathit{h}}_{\text{ud},i}^{\text{AF}}+\sqrt{\frac{1}{\mathfrak{Q}_{\text{ud},i}} \frac{1}{Z_{\text{ud},i}+1}} \tilde{\mathit{h}}_{\text{ud},i}^{\text{AF}},
\end{equation}
and the Rician distribution channel vector between HAPS-SMBS and IoT ground devices, denoted by {\color{black} $\hat{h}_{\text{sd},i}  \in \mathbb{C}^{1 \times M}$, is as follows
 \begin{equation} \label{eq:4}
\hat{h}_{\text{sd},i}=   \sqrt{\frac{1}{\mathfrak{Q}_{\text{sd},i}} \frac{Z_{\text{sd},i}}{Z_{\text{sd},i}+1}} \bar{\mathbf{h}}_{\text{sd},i}+\sqrt{\frac{1}{\mathfrak{Q}_{\text{sd},i}} \frac{1}{Z_{\text{sd},i}+1}} \tilde{\mathbf{h}}_{\text{sd},i},
\end{equation}}
where $Z_{\text{su}}$, $Z_{\text{ud},i}$, and $Z_{\text{sd},i}$ represents the Rician factors, $\mathfrak{Q}_{\text{su}}$, $\mathfrak{Q}_{\text{ud},i}$, and $\mathfrak{Q}_{\text{sd},i}$ are the large-scale path-loss coefficient between the HAPS-UAV, UAV-IoT ground devices, and HAPS-IoT ground devices, respectively. $\bar{\mathbf{h}}_{\text{su}}^{\text{AF}}  \in \mathbb{C}^{M \times 1}$, $\bar{\mathit{h}}_{\text{ud},i}^{\text{AF}}  \in \mathbb{C}^{1 \times 1}$, and $\bar{\mathbf{h}}_{\text{sd},i}  \in \mathbb{C}^{M \times 1}$ are the line of sight (LoS) components of the corresponding channels, $\tilde{\mathbf{h}}_{\text{su}}^{\text{AF}}  \in \mathbb{C}^{M \times 1}$, $\tilde{\mathit{h}}_{\text{ud},i}^{\text{AF}}  \in \mathbb{C}^{1 \times 1}$, and $\tilde{\mathbf{h}}_{\text{sd},i} \in \mathbb{C}^{M \times 1}$ are the non-line of sight (NLoS) components of the corresponding channels {\color{black}whose elements follow $\mathcal{C N} (0, 1)$}. 
We assume that the antenna at HAPS-SMBS is selected for transmission (called TAS) to find the best channel condition of the HAPS-SMBS $\rightarrow$ UAV and HAPS-SMBS $\rightarrow$ IoT ground devices links as in \cite{tran2019secure}, which are given 
 \begin{equation} \label{eq:4}
\breve{h}_{\text{su}}^{\text{AF}} =\operatorname*{arg\,max}_{1 \leq j \leq M} \mathbf{h}_{\text{su}}^{\text{AF}}, 
\end{equation}
and
{\color{black} \begin{equation} \label{eq:4}
\breve{h}_{\text{sd},i} =\operatorname*{arg\,max}_{1 \leq j \leq M} \hat{h}_{\text{sd},i}. 
\end{equation}}
IoT ground devices detect their signals based on power allocation distribution. Signal power levels are sorted in descending order, and detection follows this order. With the presence of HWI, the signal-to-interference-plus-noise ratio (SINR) for IoT ground devices is given by
\begin{equation}\label{eq:3.03}
\gamma_{\text{AF},s_{i}}= \frac{\alpha_{1} ( \Lambda^2  {P_{\text{u}}} {P_{\text{t}}} |\breve{h}_{\text{su}}^{\text{AF}}|^2| \mathit{h}_{\text{ud},i}^{\text{AF}}|^2+ |\breve{h}_{\text{sd},i}|^2 {P_{\text{t}}})}{\mathfrak{C}_{\text{a}} + \Delta_{\text{a}} + \Delta_{\text{s}} +(1 +|\mathit{h}_{\text{ud},i}^{\text{AF}}|^2 \Lambda^2  {P_{\text{u}}}) \sigma^2
} ,
\end{equation}
where \begin{math}\Delta_{\text{a}}=|\breve{h}_{\text{su}}^{\text{AF}}|^2 |\mathit{h}_{\text{ud},i}^{\text{AF}}|^2 \Lambda^2  {P}_{\text{t}} {P_{\text{u}}}  {k}_{\text{su}}^2  
+ |\mathit{h}_{\text{ud},i}^{\text{AF}}|^2 {k}_{\text{ud},i}^2 {P}_{\text{u}}  + |\breve{h}_{\text{\text{sd},i}}|^2 {k}_{\text{sd},i}^2 {P}_{\text{t}}
\end{math},  $\mathfrak{C}_{a}=\sum_{i \neq 1}^{L} \alpha_{i} ( \Lambda^2  {P_{\text{u}}} {P_{\text{t}}} |\breve{h}_{\text{su}}^{\text{AF}}|^2| \mathit{h}_{\text{ud},i}^{\text{AF}}|^2+|\breve{h}_{\text{sd},i}|^2 {P_{\text{t}}})$, and $\Delta_{\text{s}}=\Delta\mathbf{h}_{\text{sd},i}  {P_{\text{t}}} (1+{k}_{\text{sd},i}^2)$.

Hence, the sum rate of the IoT ground devices can then be computed by
\begin{equation}\label{eq:3.06}
C_{\text{AF},s_{i}}=\sum_{i=1}^{L} \mathrm{log}_{2} (1+ \gamma_{\text{AF},s_{i}} ) .
\end{equation}

\subsection{HAPS-SMBS and UAV Relaying with Passive/Active TRIS}
As shown in Fig. 1, the TRIS composed of $N$ elements is used in scenarios when the single antenna IoT ground devices are on the opposite side of the HAPS-SMBS (i.e., the incident signals penetrate through the TRIS elements). It is assumed that direct links are available between the HAPS-SMBS and IoT ground devices. 
A passive TRIS is made up of many passive components, each of which can forward the incoming signal with a phase shift that can be adjusted. Typically, every component in a passive TRIS is composed of a patch connected to an impedance-adjustable circuit for phase control. Passive RIS element consumes almost no direct-current power because of their passive operational setting, which is devoid of active radio-frequency components, and any associated thermal noise is typically minor \cite{zhang2022active}.
As presented in Fig. 1, the HAPS-SMBS transmits a superimposed coding signal
with different power allocation coefficients according to channel gain (i.e., $|\mathbf{H}_{\text{su}} \mathbf{\psi}_{\text{RIS}} \mathbf{h}_{\text{ud},1}^{\text{RIS}}|^2< |\mathbf{H}_{\text{su}} \mathbf{\psi}_{\text{RIS}} \mathbf{h}_{\text{ud},2}^{\text{RIS}}|^2< ...<|\mathbf{H}_{\text{su}} \mathbf{\psi}_{\text{RIS}} \mathbf{h}_{\text{ud},i}^{\text{RIS}}|^2$) to IoT ground devices through UAV-RIS. Thus, the received signal at IoT ground devices with $N$ element is given by
 \begin{equation} \label{eq:4}
 \begin{split}
& y_{\text{RIS},i}= \mathbf{H}_{\text{su}} \mathbf{\psi}_{\text{RIS}} \mathbf{h}_{\text{ud},i}^{\text{RIS}} (\sqrt{P_{\text{t}}}x_{\text{sc}}+\eta_{\text{t,su}}^{\text{RIS}}) + \mathbf{h}_{\text{sd},i} (\sqrt{P_{\text{t}}}x_{\text{sc}} \\& \ \ \ \ \ \ \ \ \  +  \eta_{\text{t,sd},i}^{\text{RIS}})  +\eta_{\text{r,su}}^{\text{RIS}}+\eta_{\text{r,sd},i}^{\text{RIS}} + n_{\text{r}} \mathbf{\psi}_{\text{RIS}} \mathbf{h}_{\text{ud},i}^{\text{RIS}}+\aleph,
 \end{split}
\end{equation}
where $n_{r} \sim \mathcal{CN}(0, \sigma_{r}^2)$ is the thermal noise introduced by active TRIS components, \begin{math}\mathbf{\psi}_{\text{RIS}}= \text{diag}\left(\rho_1 e^{j\theta_1}, \rho_2 e^{j\theta_2},..., \rho_N e^{j\theta_N}  \right)\end{math} is a diagonal phase-shifting matrix that fully captures the characteristics of the TRIS, $\theta_N$ is the phase shift of the $n$-th TRIS reflecting element.
The Rician distribution channel vector between the HAPS-SMBS and UAV-TRIS, denoted by $\mathbf{H}_{\text{su}}  \in \mathbb{C}^{N \times M}$, can be given as
 \begin{equation} \label{eq:4}
\mathbf{H}_{\text{su}}=   \sqrt{\frac{1}{\mathfrak{Q}_{\text{su}}} \frac{Z_{\text{su}}}{Z_{\text{su}}+1}} \bar{\mathbf{H}}_{\text{su}}+\sqrt{\frac{1}{\mathfrak{Q}_{\text{su}}} \frac{1}{Z_{\text{su}}+1}} \tilde{\mathbf{H}}_{\text{su}},
\end{equation}
and the Rician distribution channel vector between the UAV-TRIS and IoT ground devices, denoted by $\mathbf{h}_{\text{ud},i}^{\text{RIS}}  \in \mathbb{C}^{1 \times N}$, can be given as
 \begin{equation} \label{eq:4}
\mathbf{h}_{\text{ud},i}^{\text{RIS}}=   \sqrt{\frac{1}{\mathfrak{Q}_{\text{ud},i}} \frac{Z_{\text{ud},i}}{Z_{\text{ud},i}+1}} \bar{\mathbf{h}}_{\text{ud},i}^{\text{RIS}}+\sqrt{\frac{1}{\mathfrak{Q}_{\text{ud},i}} \frac{1}{Z_{\text{ud},i}+1}} \tilde{\mathbf{h}}_{\text{ud},i}^{\text{RIS}},
\end{equation}
{\color{black}in which $\bar{\mathbf{H}}_{\text{su}}  \in \mathbb{C}^{M \times N}$ and $\bar{\mathbf{h}}_{\text{ud},i}^{\text{RIS}}  \in \mathbb{C}^{N \times 1}$, are the LoS components of the corresponding channels, and $\tilde{\mathbf{H}}_{\text{su}}  \in \mathbb{C}^{M \times N}$, and $\tilde{\mathbf{h}}_{\text{ud},i}^{\text{RIS}} \in \mathbb{C}^{N \times 1}$  are the NLoS components of the corresponding channels whose elements follow $\mathcal{C N} (0, 1)$.
} $\eta_{\text{t,su}}^{\text{RIS}}$, $\eta_{\text{t,ud},i}^{\text{RIS}}$, $\eta_{\text{r,su}}^{\text{RIS}}$, and $\eta_{\text{r,ud},i}^{\text{RIS}}$ are the independent distortion noise terms defined as \cite{beddiaf2022unified,khennoufa2024error}
\begin{math}\label{eq:3}
 \mathbf {\eta}_{\text{t,su}}^{\text{RIS}} \sim \mathcal{CN}(0,\ {{k}}_{\text{t,su}}^2 {P}_{\text{t}}), 
 \end{math}
 \begin{math}\label{eq:3}
 \mathbf {\eta}_{\text{t,ud},i}^{\text{RIS}} \sim \mathcal{CN}(0, {{k}}_{\text{t,ud},i}^2 {P}_{\text{t}}), 
 \end{math}
 \begin{math}\label{eq:3}
   \mathbf {\eta}_{\text{r,su}} \sim \mathcal{CN}(0,{{k}}_{\text{r,su}}^2 {P}_{\text{t}} |\mathbf{H}_{\text{su}}|^2 ), 
 \end{math}
and
\begin{math}\label{eq:3}
\mathbf {\eta}_{\text{r,ud},i} \sim \mathcal{CN}(0, {{k}}_{\text{r,ud},i}^2 {P}_{\text{t}} |\mathbf{h}_{\text{ud},i}^{\text{RIS}} \mathbf{\psi}_{\text{RIS}}|^2).
 \end{math}
We assume $\rho_{N} = \rho$; hence, we have $\rho \leq 1$ for passive RIS, whereas $\rho > 1$ for active RIS because of its amplifiers. We assume that the TAS is used at HAPS-SMBS to find the best channel condition of the HAPS-SMBS $\rightarrow$ UAV link as given by
 \begin{equation} \label{eq:4}
\breve{\mathbf{h}}_{\text{su}} =\operatorname*{arg\,max}_{1 \leq j \leq M} \textbf{H}_{\text{su}}.
\end{equation}

The IoT ground devices detect their signals based on power allocation distribution, starting with the strongest and then the next in sequence. Assuming that the HWI exists, the SINR at IoT ground devices is given by
\begin{equation}\label{eq:3.03}
\gamma_{\text{TRIS},s_{i}}= \frac{\alpha_1(|\breve{\mathbf{h}}_{\text{su}} \mathbf{\psi}_{\text{RIS}} \mathbf{h}_{\text{ud},i}^{\text{RIS}}|^2 P_{\text{t}} + |\breve{h}_{\text{sd},i}|^2 P_{\text{t}})}{\mathfrak{C}_{\text{b}} + \sigma_{\text{r}}^2 |\mathbf{\psi}_{\text{RIS}}\mathit{\mathbf{h}}_{\text{ud},i}^{\text{RIS}}|^2  + \Delta_{\text{b}} + \Delta_{\text{s}} +\sigma^2
},
\end{equation}
where \begin{math}\Delta_{\text{b}}=|\breve{\mathbf{h}}_{\text{su}} \mathbf{\psi}_{\text{RIS}} \mathbf{h}_{\text{ud},i}^{\text{RIS}}|^2 P_{\text{t}} {k}_{\text{su}}^2+P_{\text{t}}|\breve{h}_{\text{sd},i}|^2 {k}_{\text{sd},i}^2
\end{math} and $\mathfrak{C}_{\text{b}}=\sum_{i \neq 1}^{L} \alpha_{i}(|\breve{\mathbf{h}}_{\text{su}} \mathbf{\psi}_{\text{RIS}} \mathbf{h}_{\text{ud},i}^{\text{RIS}}|^2 P_{\text{t}} + |\breve{h}_{\text{sd},i}|^2 P_{\text{t}})$.

The sum rate of the IoT ground devices is given by
\begin{equation}\label{eq:3.06}
C_{\text{TRIS},s_{i}}= \sum_{i=1}^{L} \mathrm{log}_{2} (1+ \gamma_{\text{TRIS},s_{i}} ) .
\end{equation}

\subsection{3GPP Path-loss Model}
Since platforms operate at varying altitudes and may thus encounter different attenuation phenomena, we employ realistic path-loss models as specified by 3GPP \cite{3gpp2018study,alfattani2021link}, where path-loss of the links from HAPS-SMBS to UAV, HAPS-SMBS to IoT ground devices, and UAV to IoT ground devices are defined. According to their relatively low altitude, UAV are typically taken to have full LoS conditions, thereby, the path-loss in an urban environment can be given as in \cite{3gpp2018study,alfattani2021link} by 
\begin{equation}\label{eq:3.03}
\begin{split}
&
\mathfrak{Q}_{\text{ud},i}= 28+22 \ \mathrm{log}_{10}(\mathfrak{d}_{\text{ud},i}) +20 \ \mathrm{log}_{10}(f) +\chi_{\text{ud},i},
\end{split}
\end{equation}
where $\chi_{\text{ud},i}$ is the log-normal distribution with a standard deviation given by 
\begin{math}\label{eq:6}
   \sigma_{\text{ud},i}=4.64 \ e^{-0.0066 AL_{\text{ud},i} }
\end{math}, $\mathfrak{d}_{\text{ud},i}$ is the distance between UAV and IoT ground devices, and $AL_{\text{ud},i}$ is the attitude of UAV.

Additionally, the path-loss of HAPS-SMBS for the conditions of LoS and NLoS, for the link HAPS-SMBS $\rightarrow$ UAV, can be given as in \cite{3gpp2018study,alfattani2021link} by
\begin{equation}\label{eq:6}
   \mathfrak{Q}_{\text{su}}=\mathcal{A}_{\text{su}}^{\text{LoS}} \zeta_{\text{su}}^{\text{LoS}}+\mathcal{A}_{\text{su}}^{\text{NLoS}} \zeta_{\text{su}}^{\text{NLoS}}, 
\end{equation}
and for HAPS-SMBS $\rightarrow$ IoT ground devices link is given by
\begin{equation}\label{eq:6}
   \mathfrak{Q}_{\text{sd},i}=\mathcal{A}_{\text{sd},i}^{\text{LoS}} \zeta_{\text{sd},i}^{\text{LoS}}+\mathcal{A}_{\text{sd},i}^{\text{NLoS}} \zeta_{\text{sd},i}^{\text{NLoS}}, 
\end{equation}
where 
\begin{math}
\mathcal{A}_{\text{su}}^{\iota}=b_{1} \vartheta_1^{b_{2}}+b_{3},
\end{math}
\begin{math}
\mathcal{A}_{\text{sd},i}^{\iota}=b_{1} \vartheta_2^{b_{2}}+b_{3},
\end{math}
\begin{math}
\zeta_{\text{su}}^{\iota}=\varpi_{\text{su}}^{\iota}+\mathcal{J},
\end{math}
\begin{math}
\zeta_{\text{sd},i}^{\iota}=\varpi_{\text{sd},i}^{\iota}+\mathcal{J},
\end{math}
\begin{math}
\mathcal{J}=\zeta_{\text{g}}+\zeta_{\text{s}}+\zeta_{\text{e}},
\end{math}
\begin{math}
\varpi_{\text{su}}^{\iota}=\mathfrak{L}_{\text{su}}+CL_{\text{su}}^{\iota}+\chi^{\iota},
\end{math}
\begin{math}
\varpi_{\text{sd},i}^{\iota}=\mathfrak{L}_{\text{sd},i}+CL_{\text{sd},i}^{\iota}+\chi^{\iota},
\end{math}
\begin{math}
\mathfrak{L}_{\text{su}}=32.45+20 \ \mathrm{log}_{10}(f)+20 \ \mathrm{log}_{10}(\mathfrak{d}_{\text{su}}),
\end{math}
\begin{math}
\mathfrak{L}_{\text{sd},i}=32.45+20 \ \mathrm{log}_{10}(f)+20 \ \mathrm{log}_{10}(\mathfrak{d}_{\text{sd},i}),
\end{math}
$\iota=\{\text{LoS, NLoS}\}$, $b_{1}$, $b_{2}$, and $b_{3}$ are the parameters that depend on the environment, which are determined in \cite{alfattani2021link}, $\mathfrak{d}_{\text{sd},i}$ and $\mathfrak{d}_{\text{su}}$ are the distance between HAPS-SMBS and IoT ground devices, and between HAPS-SMBS and UAV, respectively, $\vartheta_1$ and $\vartheta_2$ are the elevation angle, $\zeta_{\text{su}}^{\iota}$ and $\zeta_{\text{sd},i}^{\iota}$ are the basic path-loss, $\zeta_{\text{g}}$ is the attenuation due to atmospheric gasses, $\zeta_{\text{s}}$ is the attenuation due to either ionospheric, $\zeta_{\text{e}}$ is the building entry loss,  $\mathfrak{L}_{\text{su}}$ and $\mathfrak{L}_{\text{sd},i}$ are the free-space propagation loss, $CL_{\text{su}}^{\iota}$ and $CL_{\text{sd},i}^{\iota}$ are the clutter loss, $\chi^{\iota}$ is the shadow fading, and $f$ is the operation frequency.

\section{Energy Efficiency}
Energy efficiency of HAPS-SMBS assisted UAV with AF and active/passive UAV-TRIS for the NOMA networks can be calculated as the sum rates divided by total power consumption. As shown in \cite{zhi2022active,amin2015opportunistic}, the energy efficiency of HAPS-SMBS assisted UAV with AF and active/passive UAV-TRIS for NOMA networks can be given as 
\begin{equation} \label{eq:4}
 \begin{split}
& E_{q}=   \frac{ C_{q,s_{i}}}{P_{q}^{o}},
 \end{split}
\end{equation}
where $q=\{\text{AF, TRIS}\}$, $o=\{\text{\text{uav}, \text{act}, \text{pas}}\}$, and $P_{q}^{o}$ is the total power consumption of UAV with AF and active/passive UAV-TRIS can be modeled as in \cite{zhi2022active,amin2015opportunistic} by
 \begin{equation} \label{eq:4}
 \begin{split}
& P_{\text{q}}^{\text{o}}=   P_{\text{t}}+ \digamma_{\text{o}}+ \sum_{i=1}^{L} P_{\text{IoTGD},i}, 
 \end{split}
\end{equation}
where $\digamma_{\text{act}}=P_{\text{RIS}}^{\text{act}}+ NP_{\text{SW}}+N P_{\text{DC}}$, $\digamma_{\text{pas}}=N P_{\text{SW}} $,
$\digamma_{\text{uav}}= P_{\text{u}}+P_{\text{c}}$, $P_{\text{RIS}}^{\text{act}}$ is the output signal power of active TRIS, $P_{\text{SW}}$ is the power consumed by the phase shift switch and control circuit in each TRIS element, $P_{\text{DC}}$ is the direct current biasing power used by each active TRIS element, $P_{\text{IoTGD},i}$ is the hardware power dissipated of the $i$-th IoT ground devices, $P_{\text{c}}$ is the constant circuit power, and $P_{\text{u}}$ is the UAV power.

\begin{figure}[]
    \centering
    \includegraphics[width=0.9\columnwidth]{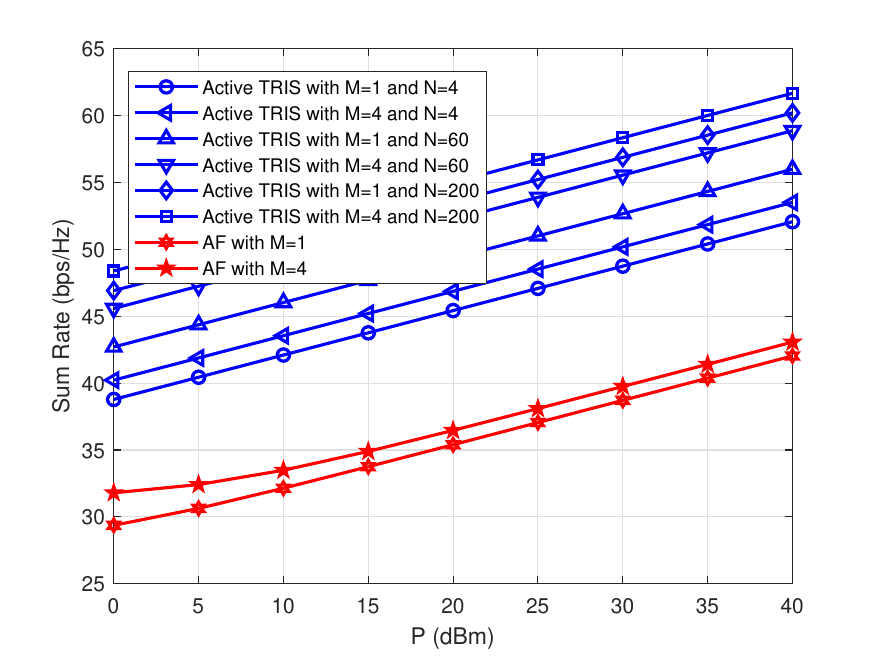}
    \caption{Sum rate w.r.t. $P$: Comparison between HAPS-SMBS with TAS assisted active UAV-TRIS and UAV using AF with the different $N$ and $M$.}
    \label{constellations}
\end{figure}

\begin{figure}[]
    \centering
      \includegraphics[width=0.9\columnwidth]{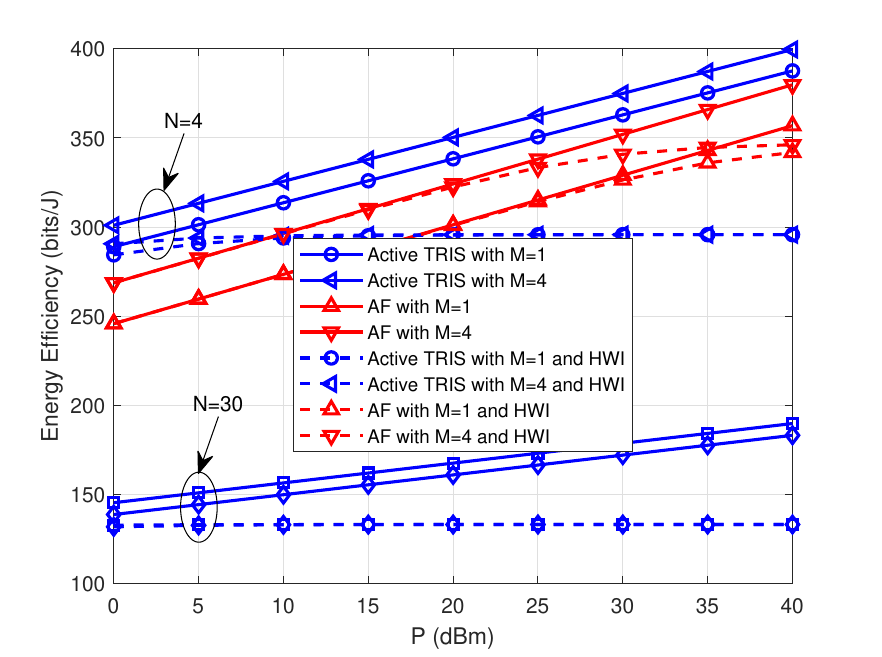}
    \caption{Energy efficiency w.r.t. $P$:  Comparison between active HAPS-SMBS with TAS assisted UAV-TRIS and UAV using AF with the different $N$ and $M$ under HWI.}
    \label{constellations}
\end{figure}

\section{Numerical Results}
In this section, the simulation results of the proposed system in terms of the sum rate and energy efficiency are obtained and compared with three schemes:  Conventional single-input single-output with HAPS-SMBS assisted UAV-TRIS, HAPS-SMBS-assisted UAV with AF supported by TAS, and HAPS-SMBS-assisted passive UAV-TRIS supported by TAS. We discuss the effect of HWI on the performance of our system. Without loss of generality, we consider that $P=P_{\text{t}}=P_{\text{u}}$, $k=K_{\text{su}}^2=K_{\text{sd},i}^2=K_{\text{ud},i}^2$, $P_{\omega}=P_{\text{DC}}=P_{\text{SW}}=P_{\text{c}}=P_{\text{RIS}}^{\text{act}}=P_{\text{RIS}}^{\text{act}}=P_{\text{IoTGD},i}$, 
 $\Xi^{LoS}=CL_{\text{su}}^{\text{LoS}}=CL_{\text{sd},i}^{\text{LoS}}$, 
 $\Xi^{\text{NLoS}}=CL_{\text{su}}^{\text{NLoS}}=CL_{\text{sd},i}^{\text{NLoS}} $, $\kappa=\zeta_{\text{g}}=\zeta_{\text{e}}$.
$k=10^{-6}$, and $P=P_{\text{t}}=P_{\text{u}}$. The parameters used in all simulations are given by $b_1=9.668$, $b_2=0.547$, $b_3=-10.58$,$L= 2$, $\Xi^{LoS}=0$ dB, $CL_{\text{su}}^{\text{NLoS}}=CL_{\text{sd},i}^{\text{NLoS}} = 14.42$ dB, $P_{\omega}=5$ dBm, $P=$ 20 dBm, $f=3$ Ghz, $\vartheta=40^{\text{o}}$, $\mathfrak{d}_{\text{ud},1}=200$ m, $\mathfrak{d}_{\text{ud},2}=100$ m, $\mathfrak{d}_{\text{sd},1}=20.2$ km, $\mathfrak{d}_{\text{sd},2}=20.1$ km, $\mathfrak{d}_{\text{su}}=20$ km, $AL_{\text{ud},i}=200$ m, $\kappa=10$ dB, and $\zeta_{\text{s}}=14.7 \vartheta^{(-1.136)}$~\cite{alfattani2021link}. {
\color{black}We assume perfect CSI in Fig. 2-5, while imperfect CSI is considered in Fig. 6.}

\begin{figure}[]
    \centering
    \includegraphics[width=0.9\columnwidth]{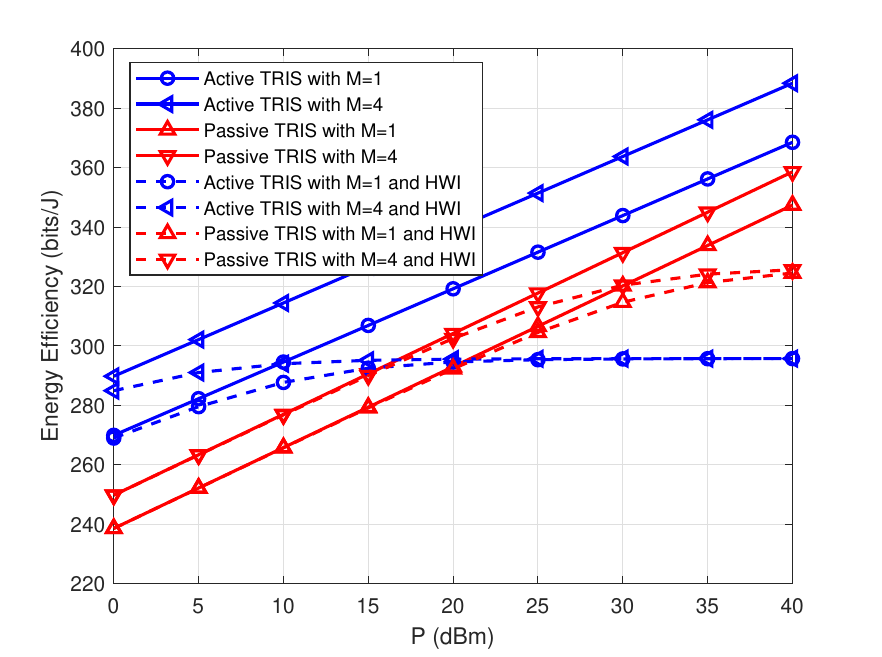}
    \caption{Energy efficiency w.r.t. $P$: Comparison between active and passive UAV-TRIS for HAPS-SMBS with TAS with the different $M$ under the effect of HWI.}
    \label{constellations}
\end{figure}

\begin{figure}[]
    \centering
      \includegraphics[width=0.9\columnwidth]{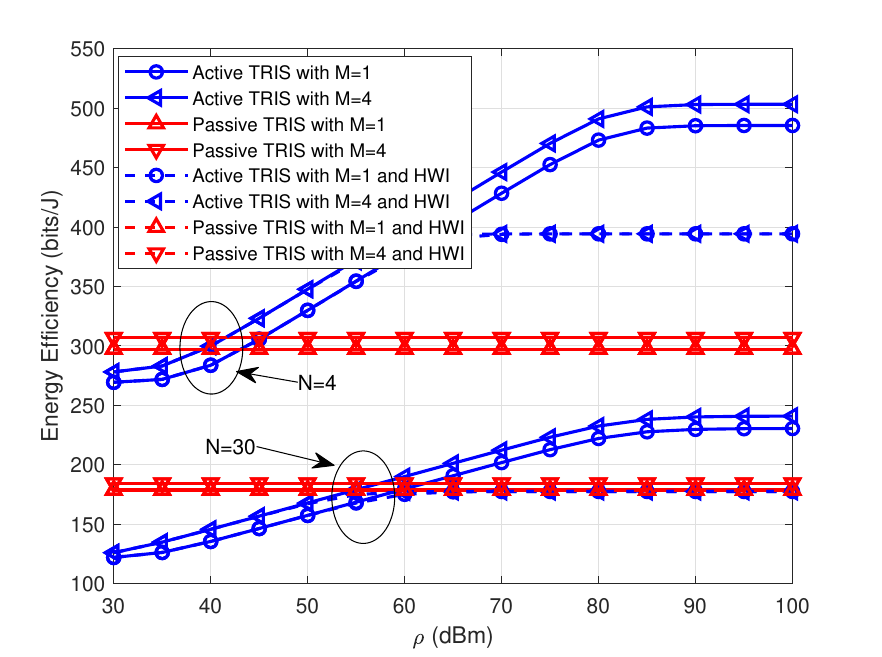}
    \caption{Energy efficiency w.r.t. $\rho$: Comparison between active and passive UAV-TRIS for HAPS-SMBS with TAS with the different $N$ and $M$ in the presence of HWI.}
    \label{constellations}
\end{figure}

In Fig. 2, we present the sum rate of HAPS-SMBS with TAS-assisted active UAV-TRIS with different numbers of TRIS elements ($N$) and antenna ($M$) in the absence of HWI. We observe that the proposed system outperforms HAPS-SMBS with TAS-assisted UAV using AF. The increasing $N$ and $M$ increase the sum rate performance. It is evident that increasing $N$ is more effective than increasing $M$ in enhancing overall performance. Increasing $N$ improves the system performance of the proposed system by around $10 \%$ compared to UAV-AF and it increases with $N$ increasing. 
To evaluate the effect of increasing $N$ on the system performance, in Fig. 3, we compare the energy efficiency of active HAPS-SMBS with TAS-assisted UAV-TRIS and HAPS-SMBS with TAS-assisted UAV using AF under the effect of HWI, when $N=\{4, 30\}$ and $M=\{1, 4\}$. We observe that the proposed system achieves higher performance gain compared to the HAPS-SMBS with TAS-assisted UAV using AF in the lower $N$ values. It is observed that increasing $N$ reduces energy efficiency performance, and this occurs due to more energy being consumed. Moreover, HWI limits the energy efficiency performance, further affected by high values of $N$, $M$, and transmit power.

\begin{figure}[]
\centering
\subfloat[]{%
\includegraphics[width=0.9\columnwidth]{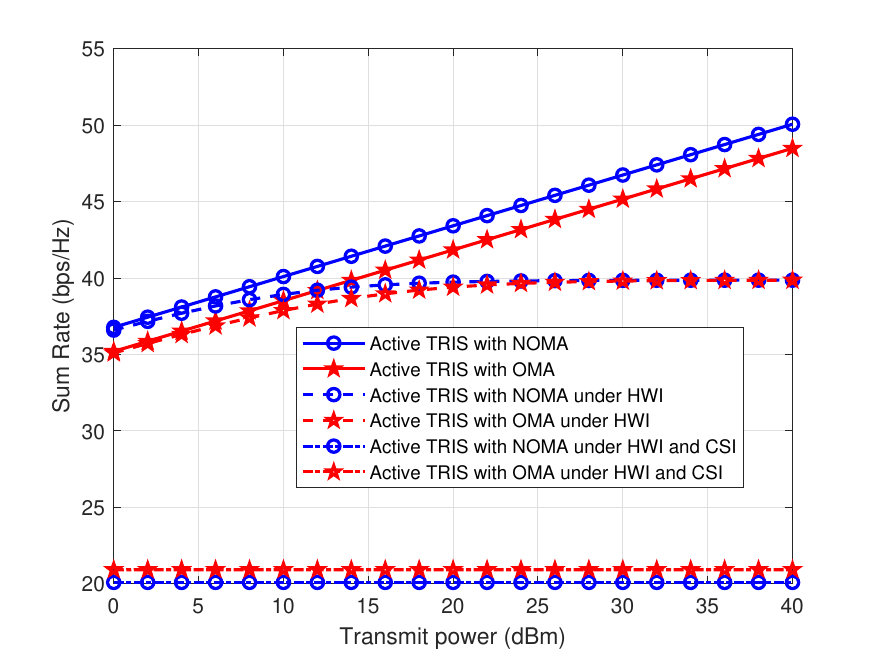}
}
\\
\subfloat[]{%
\includegraphics[width=0.9\columnwidth]{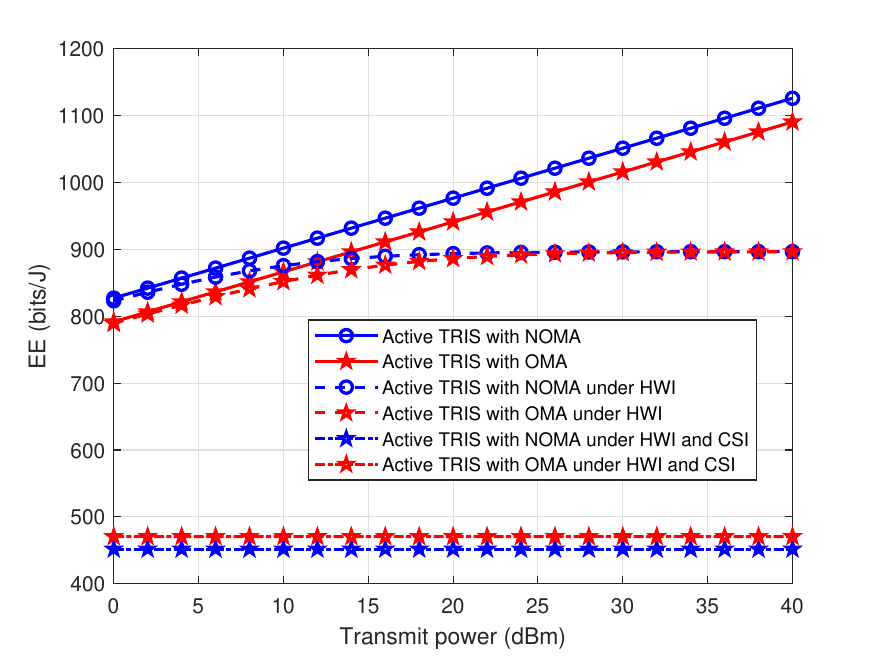}
}
\caption{{\color{black}Comparison of Sum-rate and Energy Efficiency between NOMA and OMA: (a) Sum-rate; (b) Energy Efficiency.}}
\end{figure}

The energy efficiency of the proposed system with active and passive TRIS is demonstrated in Fig. 4 with different numbers of antennas under the effect of HWI when $N=4$ and $M=\{1, 4\}$. It is seen that the active TRIS has a higher performance gain than the passive TRIS. For this reason, the signal strength received at the IoT ground devices can be greatly enhanced by active TRIS, which can use an amount of power to amplify the signal that was attenuated after the transmission in the first phase. Again, HWI has a significant impact on the performance of the system at high power levels. As we can see, HWI has a greater negative impact on the active TRIS due to the power amplification factor. 
To present the impact of the amplification factor $\rho$ on the system performance, Fig. 5 illustrates the energy efficiency of the proposed system with active and passive TRIS w.r.t. $\rho$ under the impact of HWI when $N=\{4, 30\}$ and $M=\{1, 4\}$. We observe that increasing $\rho$ improves the performance of the active TRIS compared to the passive TRIS. However, increasing $N$ reduces the energy efficiency performance in active more than passive TRIS due to the larger energy consumed. In the presence of HWI, on the other hand, the performance becomes limited again despite increasing $N$ and $M$. HWI affects system performance more than the thermal noise from active TRIS components.



{\color{black}Moreover, in Fig. 6 (a) and (b), we compare the sum-rate and energy efficiency of the active TRIS for NOMA and orthogonal multiple access (OMA) {\color{black}under HWI and imperfect CSI}. 
We observe that the NOMA system achieves higher performance gains compared to OMA.
We also observe a negative impact of HWI on the system performance, particularly at high SNR levels. {\color{black}It was also seen that the impact of imperfect CSI with HWI was significant, and its impact was less in OMA compared to NOMA. This is due to the effect of the SIC process. This illustrates the significant impact of imperfect CSI issues on the overall system performance.}}

\section{Conclusion}
In this paper, we proposed integrating active TRIS on UAVs to enhance the capabilities of a HAPS-SMBS-enabled TAS system. This approach is designed to serve IoT ground devices employing NOMA technology.
For a practical scenario, the presence of HWI {\color{black}and imperfect CSI} was also taken into account in our proposed system, and thus the sum rate and energy efficiency of the proposed system were obtained under the effect of HWI {\color{black}and imperfect CSI}. 
As a benchmark, the proposed system was compared with three schemes: Conventional single-input single-output system with HAPS-SMBS assisted UAV-TRIS, HAPS-SMBS enabled by TAS-assisted UAV relaying without TRIS, and HAPS-SMBS with TAS-assisted passive UAV-TRIS. 
The results revealed that the proposed system achieves a higher performance gain compared to the benchmarks. 
Moreover, the increasing number of TRIS elements increases the sum rate performance; however, it consumes a higher power compared to passive TRIS.
HWI {\color{black}and imperfect CSI} limited the system performance at the high transmit power levels despite the increased number of TRIS elements.

%




\section*{Acknowledgment}
This study is supported in part by TÜBİTAK (The Scientific and Technological Research Council of Türkiye), in part by the Study in Canada Scholarship (SICS) by Global Affairs Canada, in part by the Ministry of Higher Education and Scientific Research of Algeria (MESRS), and in part by the Discovery Grant RGPIN-2022- 05231 from the Natural Sciences and Engineering Research Council of Canada (NSERC).


\bibliographystyle{ieeetr}

\end{document}